\documentclass[%
 aip,
 amsmath,amssymb,
 reprint,%
]{revtex4-1}

\usepackage{graphicx}% Include figure files
\usepackage{dcolumn}% Align table columns on decimal point
\usepackage{bm}% bold math
\usepackage[utf8]{inputenc}
\usepackage[T1]{fontenc}
\usepackage{mathptmx}
\usepackage{epstopdf,mathrsfs}
\usepackage{multirow}
\begin{document}

\preprint{AIP/123-QED}

\title[]{Different effects of external force fields on aging L\'{e}vy walk}
% Force line breaks with \\

\author{Yao Chen}
\email{ychen@njau.edu.cn}
 \affiliation{College of Sciences, Nanjing Agricultural University, Nanjing, 210095, P.R. China}%Lines break automatically or can be forced with \\
%\author{}%
%\affiliation{
%$^2$School of Mathematics and Statistics, Nanjing University of Science and Technology, Nanjing, 210094, P.R. China
%}%

\author{Xudong Wang}
 %\homepage{http://www.Second.institution.edu/~Charlie.Author.}
\affiliation{%
School of Mathematics and Statistics, Nanjing University of Science and Technology, Nanjing, 210094, P.R. China
}%


\begin{abstract}
Aging phenomena have been observed in numerous physical systems. Many statistical quantities depend on the aging time $t_a$ for aging anomalous diffusion processes. This paper pays more attention to how an external force field affects the aging L\'{e}vy walk.
Based on the Langevin picture of L\'{e}vy walk and generalized Green-Kubo formula, we investigate the quantities which include the ensemble- and time-averaged mean-squared displacements in both weak aging $t_a\ll t$ and strong aging $t_a\gg t$ cases, and compare them to the quantities in the absence of any force field. Two typical force fields, constant force $F$ and time-dependent periodic force $F(t)=f_0\sin(\omega t)$, are considered for comparison. The generalized Einstein relation is also discussed in the case with constant force.
We find that the constant force is the key of generating the aging phenomena and enhancing the diffusion behavior of aging L\'{e}vy walk, while the time-dependent periodic force is not. The different effects of the two kinds of forces on the aging phenomena of L\'{e}vy walk are verified by both theoretical analyses and numerical simulations.
\end{abstract}

\maketitle

\begin{quotation}
Aging phenomena are common in many complex dynamical systems, where the dynamical properties of the focused system depend on the aging time $t_a$ even in
the limit that the measurement time is sufficiently long. Based on a typical anomalous diffusion model, L\'{e}vy walk, this paper focuses on the effects of the external force fields on aging processes. The difficulty of dealing with the time-space coupling of L\'{e}vy walk can be avoided by using the Langevin picture of L\'{e}vy walk in an external force field, together with the method of the generalized Green-Kubo formula. By comparing the two typical force fields, constant force $F$ and time-dependent periodic force $F(t)=f_0\sin(\omega t)$, we find that the external forces play different roles in aging phenomena.

\end{quotation}

\section{Introduction}

Due to the complexity of the transport phenomena in the nature, a common assumption is made in the analysis  that the motion of particles  commences at the moment of observation. That is, the observation begins immediately after the preparation of the system. However, in a multitude of experimental setups, we cannot start to observe the system at the beginning of its evolution. Assume the measurement starts at some time $t_a >0$ after the initiation of the process at $t=0$. {The time $t_a$ is called age}.
In many cases the delay $t_a$ largely changes the statistical properties of the observed process. Such a phenomenon is called aging, a term
which was originally used in the area of glassy materials \cite{MonthusBouchaud:1996,BertinBouchaud:2003,BurovBarkai:2007}. A system exhibits aging if its dynamical properties depend on the aging time $t_a$ even in the limit that the measurement time is sufficiently long. Aging phenomena have been found in many complex dynamical systems like spin glasses, glasses, polymers \cite{Struick:1978}, and in random
walks in random environments \cite{LalouxDoussal:1998}.

L\'{e}vy walk is one of the typical models to describe superdiffusion phenomena in the nature \cite{SanchoLacastaLindenbergSokolovRomero:2004,ZaburdaevDenisovHanggi:2013,RebenshtokDenisovHanggiBarkai:2014,ZaburdaevDenisovKlafter:2015}. This model is characterized by coupled continuous time random walk (CTRW) \cite{ShlesingerKlafterWong:1982,KlafterBlumenShlesinger:1987,Zaburdaev:2006,ZaburdaevDenisovKlafter:2015}, where the waiting time and jump length are coupled through a constant velocity. Taking different values of the power-law exponent $\alpha$ of the distribution of each unidirection flight time, L\'{e}vy walk could describe ballistic diffusion ($0<\alpha<1$), sub-ballistic superdiffusion ($1<\alpha<2$) and normal diffusion ($\alpha>2$).
L\'{e}vy walk has been successfully applied in various fields, including the electron transfer \cite{Nelson:1999}, dispersion in turbulent systems \cite{SolomonWeeksSwinney:1993}, the anomalous superdiffusion of cold atoms in optical lattices \cite{KesslerBarkai:2012}, endosomal active transport within living cells \cite{ChenWangGranick:2015}, and so on.

The aging CTRW is introduced by Monthus and Bouchaud \cite{MonthusBouchaud:1996} as a simple phenomenological model of aging dynamics in glasses, and its biased and nonbiased versions are investigated by using fractal renewal theory \cite{BarkaiCheng:2003}.
Aging ballistic L\'{e}vy walk with $0<\alpha<1$ has been studied in the context of ensemble-averaged mean-squared displacement (EAMSD), time-averaged mean-squared displacement (TAMSD), as well as the disparity between these two quantities and its relation to ergodicity breaking  \cite{MagdziarzZorawik:2017}. The aging phenomena have also been found in other anomalous diffusion processes \cite{DechantLutzKesslerBarkai:2014,SchulzBarkaiMetzler:2014,Stage:2017,SongMoonJeonPark:2018,WangChenDeng:2019-2}.
Nevertheless, the aging L\'{e}vy walk with different $\alpha$, especially in the presence of an external force field, has not been investigated. This paper focuses on the effects of the external force fields on aging L\'{e}vy walk by considering two typical forces, constant force $F$ and time-dependent periodic force $F(t)=f_0\sin(\omega t)$. The constant force acting on non-aging anomalous diffusion processes has been discussed a lot \cite{MetzlerKlafter:2000,BelBarkai:2005_2,EuleFriedrich:2009,AkimotoCherstvyMetzler:2018,ChenWangDeng:2019-2,ChenWangDeng:2019-3}, and the case of time-dependent periodic force has also been investigated in Refs. \cite{SokolovKlafter:2006,MagdziarzWeronKlafter:2008,ChenWangDeng:2019-2,ChenDeng:2021}.

The main focused quantities of this paper are the aging EAMSD and TAMSD, which are defined by, for a biased process with {age time $t_a$},
\begin{equation}\label{EAdefination}
\begin{split}
\langle \Delta x^2_{t_a}(t)\rangle
&:=\langle x_{t_a}^2(t)\rangle-\langle x_{t_a}(t)\rangle^2\\
&=\langle (x(t_a+t)-x(t_a))^2\rangle-\langle x(t_a+t)-x(t_a)\rangle^2,
\end{split}
\end{equation}
and
\begin{equation}\label{TAdefination}
\begin{split}
  \overline{\delta_{t_a}^2(\Delta)}
  &=\frac{1}{T-\Delta}\int_{t_a}^{t_a+T-\Delta}  [x(t+\Delta)-x(t)  \\
  &~~~    -\langle x(t+\Delta)-x(t)\rangle]^2dt,
\end{split}
\end{equation}
respectively. Here, the $\Delta$ in Eq. \eqref{EAdefination} denotes the subtracting the aging first moment for a biased process, while that in Eq. \eqref{TAdefination} represents the lag time and separates the displacement between trajectory points. The latter one should be much shorter than measurement time $T$ to obtain good statistical properties, i.e., $\Delta\ll T$. For a non-biased process, the first moments in Eqs. \eqref{EAdefination} and \eqref{TAdefination} vanish and these definitions recover the simple forms \cite{MetzlerJeonCherstvyBarkai:2014,MagdziarzZorawik:2017}. The TAMSD is common used to analyze the particle's trajectory in single particle tracking experiments, which have been widely employed to study the diffusion of particles in living cell \cite{GoldingCox:2006,WeberSpakowitzTheriot:2010,BronsteinIsraelKeptenMaiTalBarkaiGarini:2009}.

Considering the significant advantage of Langevin equation when describing the anomalous processes influenced by the external force fields, we conduct the investigations on aging L\'{e}vy walk based on a set of Langevin equations coupled with a subordinator. The equivalence between the Langevin picture and the CTRW form of L\'{e}vy walk has been presented in Refs. \cite{WangChenDeng:2019,ChenWangDeng:2019-3,EuleZaburdaevFriedrichGeisel:2012}.
Then we analyze the moments, EAMSD and TAMSD, and compare them to those in the absence of any force field to reflect the different effects of constant force and time-dependent periodic force on the aging L\'{e}vy walk. All the results we obtain can recover to the non-aging case by taking $t_a=0$, and recover to the force-free case by taking the external force $F=0$.

In the case of the aging L\'{e}vy walk affected by the constant force $F$, we also check the effectiveness of the generalized Einstein relation \cite{MetzlerBarkaiKlafter:1999,BarkaiMetzlerKlafter:2000,MetzlerKlafter:2000,BlickleSpeckLutzSeifertBechinger:2007}, which connects the first moment of the particle displacement under a constant force to the second moment of the free particle. The generalized Einstein relation is valid for non-aging L\'{e}vy walk with respect to the EAMSD, but failed with respect to the TAMSD \cite{FroembergBarkai:2013,FroembergBarkai:2013-3,ChenWangDeng:2019-3}.

The structure of this paper is as follows. In Sec. \ref{two}, we show the scaling Green-Kubo formula, which gives the general expressions of EAMSD and ensemble-averaged TAMSD of the aging stochastic process by using the asymptotic scaling form of velocity correlation function. In Sec. \ref{three} and Sec. \ref{four}, we analyze the moments, EAMSD and ensemble-averaged TAMSD in the case of the constant force and time-dependent periodic force, respectively. Finally, we give some summaries in Sec. \ref{five}.

\section{Generalized Green-Kubo formula}\label{two}

The Green-Kubo formula is a central result of the nonequilibrium statistical mechanics, which relates the spatial diffusion coefficient $D$ of the system to the integral of the stationary velocity correlation function $\langle v(t)v(t+\tau)\rangle$ \cite{Taylor:1922,Green:1954,Kubo:1957}. The velocity correlation function of Brownian motion decays exponentially with respect to the lag time $\tau$, and thus, it is integrable over the entire timeline. For various anomalous diffusion processes, the velocity correlation functions might be nonstationary and decay in a power-law rate, which leads to the divergence of the integral and the failure of the classical Green-Kubo formula. Therefore, a generalized Green-Kubo formula was proposed in Ref. \cite{DechantLutzKesslerBarkai:2014}, which reveals the relation between the diffusion properties of the stochastic process and the nonstationary velocity correlation function. This formula enables the direct evaluation of the EAMSD from the knowledge of the scaling properties of the velocity correlation function. It is also extended to evaluate the ensemble-averaged TAMSD by using the velocity correlation function for many velocity-jump processes, such as L\'{e}vy walk and its variants \cite{MeyerBarkaiKantz:2017,WangChenDeng:2019,ChenWangDeng:2019-3,ChenDeng:2021}.
Here we show the main results of the generalized Green-Kubo formula, i.e., the relation between the MSDs of the concerned diffusion process and the nonstationary velocity correlation function, which is also valid for aging processes.

Considering the stochastic process whose velocity correlation function has the asymptotic scaling form for large $t$ and large lag time $\tau$,
\begin{equation}\label{vv}
\begin{split}
\langle v(t)v(t+\tau)\rangle\simeq Ct^{\nu-2}\phi\left(\frac{\tau}{t}\right),
\end{split}
\end{equation}
where $C$ is a positive constant and $\phi(s)$ is a positive-valued scaling function when $s\rightarrow0$ and $s\rightarrow\infty$.
Then the aging second moment of process $x(t)$ is given by \cite{DechantLutzKesslerBarkai:2014,MeyerBarkaiKantz:2017}
\begin{equation}\label{x2}
\begin{split}
\langle  x^2_{t_a}(t)\rangle&=\langle (x(t_a+t)-x(t_a))^2\rangle\\
                                 &=\int_{t_a}^{t_a+t}\int_{t_a}^{t_a+t} \langle v(t_1)v(t_2)\rangle dt_1dt_2\\
                                 &\simeq 2D_\nu ^{t/t_a}t^\nu,
\end{split}
\end{equation}
where the last line is obtained by substituting Eq. \eqref{vv} into the integrand together with a variable substitution. If $x(t)$ has null mean value, then Eq. \eqref{x2} is equal to the aging EAMSD.
The diffusion coefficient is
\begin{equation}
\begin{split}
D_\nu ^{t/t_a}&=C \int_0^1 z^{\nu-1}\left(1+\frac{1}{z\frac{t}{t_a}}\right)^{\nu-1} \\
             &~~~~\times\int_0^{z\frac{t}{t_a}}(s+1)^{-\nu}\phi(s)dsdz,
\end{split}
\end{equation}
depending on both the observation time $t$ and {aging time $t_a$}. It tends to a constant as $t\rightarrow\infty$ for weak aging case with $t\gg t_a$ \cite{DechantLutzKesslerBarkai:2014,MeyerBarkaiKantz:2017}, i.e.,
\begin{equation}\label{4}
\begin{split}
D_\nu ^{\infty}=D_\nu =\frac{C}{\nu}\int_0^\infty (s+1)^{-\nu}\phi(s)ds.
\end{split}
\end{equation}
The above result is independent of the {aging time $t_a$},  which is same as the non-aging case. On the other hand, for strong aging case with $t\ll t_a$, one has the asymptotic form \cite{DechantLutzKesslerBarkai:2014,MeyerBarkaiKantz:2017}
\begin{equation}\label{5}
\begin{split}
D_\nu ^{t/t_a}\simeq \frac{cC}{(\nu-\beta-1)(\nu-\beta)}\left(\frac{t_a}{t}\right)^\beta,
\end{split}
\end{equation}
where the constants $\beta$ and $c$ come from the velocity's variance $\langle v^2(t)\rangle\propto t^\beta$ and the asymptotic form of $\phi(s)$ for small $s$: $\phi(s)\simeq cs^{-\delta}$, respectively. There also exists an intrinsic relevance $\delta=2-\nu+\beta$ between the exponents. In contrast to Eq. \eqref{4}, the dependence of the diffusion coefficient on aging time $t_a$ in Eq. \eqref{5} implies the aging behavior in strong aging case. In this case, the dependence on $t_a$ and the aging behavior  vanish only when the velocity process tends to a stationary steady state with constant variance ($\beta=0$). Observing the expressions of diffusion coefficients in Eqs. \eqref{4} and \eqref{5}, we also find that whatever $\beta=0$ or not, only the small-$s$ asymptotics of $\phi(s)$ is effective in the strong aging case, but the entire behavior of $\phi(s)$ is needed in the weak aging case. This is the essential reason of the discrepancy between aging and non-aging EAMSDs.

Nevertheless, based on the definition of TAMSD in Eq. \eqref{TAdefination} and the priori condition $\Delta\ll T$, the asymptotic behavior $\Delta\ll t$ in the integrand plays the leading role, i.e., only the small-$s$ asymptotics of $\phi(s)$ is effective. More precisely, for the ensemble-averaged TAMSD of the aging stochastic process $x(t)$ with null mean value, by virtue of Eqs. \eqref{x2} and \eqref{5}, one arrives at the asymptotic form
\begin{equation}\label{TA}
\begin{split}
\langle \overline{\delta_{t_a}^2(\Delta)}\rangle&=\frac{1}{T-\Delta}\int_{t_a}^{t_a+T-\Delta}
\langle (x(t+\Delta)-x(t))^2\rangle dt\\
&\simeq
                     \left\{
    \begin{array}{ll}
    \frac{2cC}{(\beta+1)(\nu-\beta-1)(\nu-\beta)}T^\beta\Delta^{\nu-\beta}, & T\gg t_a, \\[4pt]
    \frac{2cC}{(\nu-\beta-1)(\nu-\beta)}t_a^\beta\Delta^{\nu-\beta}, & T\ll t_a. \\[4pt]
\end{array}
  \right.
\end{split}
\end{equation}
The key parameters $\nu$ and $\beta$ come from the large-$t$ behavior of the velocity correlation function and {the small-$s$ behavior of $\phi(s)$} in Eq. \eqref{vv}, respectively.

If the variance of the velocity process tends to a constant at long time limit, i.e., $\beta=0$, then the ensemble-averaged TAMSD shows the independence of the aging time $t_a$ and measurement time $T$, and the two asymptotics in Eq. \eqref{TA} are equal to each other. The typical examples include the classical Brownian motion and L\'{e}vy walk, representing normal diffusion and superdiffusion, respectively. In detail, based on the generalized Green-Kubo formula, the L\'{e}vy walk with $0<\alpha<1$ has the EAMSD
\begin{equation}\label{freeEAMSD1}
\begin{split}
\langle  x_{t_a}^2(t)\rangle_0
&\simeq
                     \left\{
    \begin{array}{ll}
    \frac{D}{\gamma}(1-\alpha)t^2, & t\gg t_a, \\[4pt]
    \frac{D}{\gamma}t^2, & t\ll t_a, \\[4pt]
\end{array}
  \right.
\end{split}
\end{equation}
and the ensemble-averaged TAMSD
\begin{equation}\label{freeTAMSD1}
\begin{split}
\langle \overline{\delta_{t_a}^2(\Delta)}\rangle_0
&\simeq
    \frac{D}{\gamma}\Delta^2
\end{split}
\end{equation}
for both $T\gg t_a$ and $T\ll t_a$.  While for $1<\alpha<2$,  the EAMSD is
\begin{equation}\label{freeEAMSD2}
\begin{split}
\langle x_{t_a}^2(t)\rangle_0
&\simeq
                     \left\{
    \begin{array}{ll}
    \frac{2D(\alpha-1)}{\gamma(3-\alpha)(2-\alpha)}t^{3-\alpha}, & t\gg t_a, \\[4pt]
    \frac{2D}{\gamma(3-\alpha)(2-\alpha)}t^{3-\alpha}, & t\ll t_a, \\[4pt]
\end{array}
  \right.
\end{split}
\end{equation}
and the ensemble-averaged TAMSD is
\begin{equation}\label{freeTAMSD2}
\begin{split}
\langle \overline{\delta_{t_a}^2(\Delta)}\rangle_0
&\simeq
    \frac{2D}{\gamma(3-\alpha)(2-\alpha)}\Delta^{3-\alpha}
\end{split}
\end{equation}
for $T\gg t_a$ and $T\ll t_a$. Note that the results in Eqs. \eqref{freeEAMSD1}-\eqref{freeTAMSD2} are presented in the form with parameters $D$ and $\gamma$, which will be introduced in the next section of this paper. The subscript ``$0$'' denotes the force-free case of aging L\'{e}vy walk.
The Eqs. \eqref{freeEAMSD1} and \eqref{freeTAMSD1} for $0<\alpha<1$ are consistent to the results in Ref. \cite{MagdziarzZorawik:2017}, where the aging ballistic L\'{e}vy walks are studied. The Eqs. \eqref{freeEAMSD1}-\eqref{freeTAMSD2}  can also be obtained by taking the external force $F=0$ in the next section. We present them here as a specific case for the application of the generalized Green-Kubo formula, and for a comparison with the MSDs in the cases with external forces.

All the results in the weak aging cases with $t\gg t_a$ and $T\gg t_a$ above are same as the non-aging case $t_a=0$. The aging has a weak effect on EAMSD in the case $t\ll t_a$ in Eqs. \eqref{freeEAMSD1} and \eqref{freeEAMSD2}, since it only increases the diffusion coefficient without changing the diffusion behavior. While for the ensemble-averaged TAMSD, it shows the same results for both weak and strong aging cases and
the aging does not yield any effects.

We have presented the procedures of evaluating the MSDs of zero mean processes by using the generalized Green-Kubo formula. For aging L\'{e}vy walk in a force field, however, the displacement is usually biased and has nonzero mean value. In this case, we use the generalized Green-Kubo formula to calculate the part of the second moment, and subtract the square of the first moment, then we get the EAMSD and ensemble-averaged TAMSD, and reveal how the external forces influence the aging L\'{e}vy walk. As two kinds of common external forces, constant force $F$ and time-dependent periodic force $F(t)=f_0\sin(\omega t)$ will be investigated separately.

\section{Aging L\'{e}vy walk in a constant force field}\label{three}

The Langevin picture of L\'{e}vy walk model affected by the constant force $F$ is given by \cite{ChenWangDeng:2019-3}
\begin{equation}\label{LW_force}
\begin{split}
    \frac{d}{d t}x(t)&=v(t),\\
    \frac{d}{d s}v(s)&=-\gamma v(s) +F \eta(s)+\xi(s),\\
    \frac{d}{d s}t(s)&= \eta(s).
\end{split}
\end{equation}
Another alternative way of investigating the L\'{e}vy walk under the constant force $F$ is based on the collision model proposed in Ref. \cite{BarkaiFleurov:1998}. For aging L\'{e}vy walk, however, the Langevin equations in Eqs. \eqref{LW_force} are more convenient.
The initial position and velocity are assumed to be $x(0)=v(0)=0$.
Here $\gamma$ is the friction coefficient, $\xi(s)$ is a Gaussian white noise with zero mean $\langle \xi(s)\rangle=0$ and correlation function $\langle \xi(s_1)\xi(s_2)\rangle=2D\delta(s_1-s_2)$. The L\'{e}vy noise $\eta(s)$, regarded as the formal derivative of the $\alpha$-dependent subordinator $t(s)$ \cite{BauleFriedrich:2005,WangChenDeng:2019}, is independent of the Gaussian white noise $\xi(s)$. The constant force $F$ is multiplied by the L\'{e}vy noise $\eta(s)$ in the second sub-equation, which implies that the constant force $F$ acts on the diffusion process throughout all physical time $t$ \cite{CairoliBaule:2015-2,ChenWangDeng:2019-2,ChenWangDeng:2019-3}. Otherwise, the effect of constant force $F$ vanishes when the inner time $s(t)$ does not change, which corresponds to the unidirection motion of L\'{e}vy walk. The derivative of position $x$ with respect to physical time $t$ is velocity $v$, and the subordinator $t(s)$ is aimed to characterize the distribution of duration of each flight of L\'{e}vy walk. When $F=0$, the Langevin equation \eqref{LW_force} recovers to the force-free case \cite{WangChenDeng:2019}.

With the help of the inverse $\alpha$-dependent subordinator $s(t):=\inf_{s>0}\{s:t(s)>t\}$ \cite{BauleFriedrich:2005}, the velocity process in physical time $t$ is defined as $v(t):=v(s(t))$, which equals to
\begin{equation}\label{LWt}
\begin{split}
v(t)&=F\int_0^t e^{-\gamma(s(t)-s(t'))}dt'  \\
&~~~+\int_0^t e^{-\gamma(s(t)-s(t'))}\xi(s(t'))ds(t').
\end{split}
\end{equation}
The first term on the right-hand side is contributed by the constant force $F$, while the another term represents the impact of surrounding environment, i.e., the random deriving force $\xi(s)$. It can be found that the moments of $v(t)$ depends on the two-point joint PDF $h(s_1,t_1;s_2,t_2)$ of the inverse subordinator $s(t)$, which is (in Laplace space $t_1\rightarrow\lambda_1,t_2\rightarrow\lambda_2$) \cite{BauleFriedrich:2005,WangChenDeng:2019}:
\begin{equation}\label{h2}
\begin{split}
&h(s_1,\lambda_1;s_2,\lambda_2)  \\
    &=\frac{\partial}{\partial s_1} \frac{\partial}{\partial s_2} \frac{1}{\lambda_1\lambda_2}\,g(\lambda_1,s_1;\lambda_2,s_2) \\
    &=\delta(s_2-s_1)\frac{\Phi(\lambda_1)+\Phi(\lambda_2)-\Phi(\lambda_1+\lambda_2)}{\lambda_1\lambda_2}\,{ e}^{-s_1\Phi(\lambda_1+\lambda_2)} \\
    &~~~+\Theta(s_2-s_1)\frac{\Phi(\lambda_2)(\Phi(\lambda_1+\lambda_2)-\Phi(\lambda_2))}{\lambda_1\lambda_2} \\
    &~~~\times{ e}^{-s_1\Phi(\lambda_1+\lambda_2)}{ e}^{-(s_2-s_1)\Phi(\lambda_2)}   \\
    &~~~+\Theta(s_1-s_2)\frac{\Phi(\lambda_1)(\Phi(\lambda_1+\lambda_2)-\Phi(\lambda_1))}{\lambda_1\lambda_2} \\
    &~~~\times{ e}^{-s_2\Phi(\lambda_1+\lambda_2)}{ e}^{-(s_1-s_2)\Phi(\lambda_1)},
\end{split}
\end{equation}
where $\Phi(\lambda)= \lambda^\alpha$ for $0<\alpha<1$ \cite{BauleFriedrich:2005} and $\Phi(\lambda)= \tau_0/(\alpha-1)\lambda-{\tau_0^\alpha} |\Gamma(1-\alpha)| \lambda^\alpha$  for $1<\alpha<2$ \cite{WangChenDeng:2019}. For convenience, we take the characteristic time $\tau_0=1$ in $\Phi(\lambda)$ in this paper. Therefore, the first moment of velocity process in physical time is \cite{ChenWangDeng:2019-3}
\begin{equation}\label{mean-v}
\begin{split}
{\langle v(t)\rangle} \simeq \left\{
    \begin{array}{ll}
    F(1-\alpha)t, &~~ 0<\alpha<1, \\[4pt]
    \frac{F(\alpha-1)}{2-\alpha}t^{2-\alpha}, &~~ 1<\alpha<2.
\end{array}
  \right.
\end{split}
\end{equation}
Performing the integral over time interval $[t_a,t_a+t]$ yields the mean position of the aging L\'{e}vy walk
\begin{equation}\label{mean1}
\begin{split}
\langle x_{t_a}(t)\rangle \simeq \left\{
    \begin{array}{ll}
       \frac{F(1-\alpha)}{2}t^2, & t\gg t_a, \\[4pt]
      F(1-\alpha)t_at, & t\ll t_a,
\end{array}
  \right.
\end{split}
\end{equation}
for $0<\alpha<1$, and
\begin{equation}\label{mean2}
\begin{split}
\langle x_{t_a}(t)\rangle \simeq \left\{
    \begin{array}{ll}
    \frac{F(\alpha-1)}{(2-\alpha)(3-\alpha)}t^{3-\alpha}, & t\gg t_a, \\[4pt]
     \frac{F(\alpha-1)}{(2-\alpha)}t_a^{2-\alpha}t, & t\ll t_a,
\end{array}
  \right.
\end{split}
\end{equation}
for $1<\alpha<2$. The weak aging cases with $t\gg t_a$ in Eqs. \eqref{mean1} and \eqref{mean2} are consistent to the non-aging case in Ref. \cite{ChenWangDeng:2019-3}, presenting superdiffusion behaviors with different diffusion exponents for different $\alpha$, while the strong aging cases with $t\ll t_a$ both yield the linear growth with respect to the observation time $t$.
For aging subdiffusive CTRW with power-law-distributed waiting time under the constant force $F$, the first moment $\langle x_{t_a}(t)\rangle$ shows the same asymptotic behavior as the EAMSD of the force-free case \cite{BarkaiCheng:2003}. However, it shows different asymptotic behaviors for aging L\'{e}vy walk between the first moment in Eqs. \eqref{mean1} and \eqref{mean2} and the EAMSD in Eqs. \eqref{freeEAMSD1} and \eqref{freeEAMSD2}.
The first moments in Eqs. \eqref{mean1} and \eqref{mean2} will also be used to calculate the MSDs in the following.

Now considering the aging Einstein relation \cite{BarkaiCheng:2003}
\begin{equation}\label{AER}
  \langle x_{t_a}(t)\rangle = \frac{\langle x_{t_a}^2(t)\rangle_0}{2k_B\mathcal{T}}F,
\end{equation}
where  $\langle x_{t_a}^2(t)\rangle_0$ and $\langle x_{t_a}(t)\rangle$ denote the EAMSD of the free aging L\'{e}vy walk without an external force and the first moment of aging L\'{e}vy walk in the presence of the constant force field $F$, respectively. The effective kinetic temperature $k_B\mathcal{T}$ is equal to $D/\gamma$ for the Langevin system in Eq. \eqref{LW_force}. Comparing the relationship between the EAMSDs $\langle x_{t_a}^2(t)\rangle_0$ in Eqs. \eqref{freeEAMSD1} and \eqref{freeEAMSD2} and the first moments $\langle x_{t_a}(t)\rangle$ in Eqs. \eqref{mean1} and \eqref{mean2}, we find that for both $0<\alpha<1$ and $1<\alpha<2$, the aging Einstein relation is satisfied by the weak aging L\'{e}vy walk, which is consistent to the non-aging case \cite{FroembergBarkai:2013-3,ChenWangDeng:2019-3}. But for the strong aging L\'{e}vy walk, the aging Einstein relation in Eq. \eqref{AER} is not satisfied due to the $t_a$-dependence in the first moment  $\langle x_{t_a}(t)\rangle$ for any $\alpha$.

Besides, no matter $0<\alpha<1$ and $1<\alpha<2$, we find that the generalized Einstein relation does not hold with respect to the time average, i.e.,
\begin{equation}\label{AER2}
\begin{split}
\langle \overline{\delta^1(\Delta)}\rangle \neq \frac{\langle\overline{\delta^2(\Delta)} \rangle_0}{2k_B \mathcal{T}}F,
\end{split}
\end{equation}
where
\begin{equation}\label{Def-firstTA}
\langle\overline{\delta^1(\Delta)}\rangle=\frac{1}{T-\Delta}\int_0^{T-\Delta}\langle x(t+\Delta)-x(t)\rangle dt
\end{equation}
denotes the first moment of the time average in the presence of the constant force and $\langle\overline{\delta^2(\Delta)} \rangle_0$ is the ensemble-averaged TAMSD of free aging L\'{e}vy walk. More exactly, when $0<\alpha<1$, taking $\langle x(t+\Delta)-x(t)\rangle\simeq F(1-\alpha)\Delta t$ in Eq. \eqref{mean1} into Eq. \eqref{Def-firstTA}, one has
 \begin{equation}
\langle\overline{\delta^1(\Delta)}\rangle=\frac{F(1-\alpha)}{2}T\Delta.
\end{equation}
Comparing it with the ensemble-averaged TAMSD of free aging L\'{e}vy walk in Eq. \eqref{freeTAMSD1}, we find the generalized Einstein relation for time average does not hold. Similarly, for the case with $1<\alpha<2$, taking $\langle x(t+\Delta)-x(t)\rangle\simeq\frac{F(\alpha-1)}{2-\alpha}\Delta t^{2-\alpha}$ in Eq. \eqref{mean2} into Eq. \eqref{Def-firstTA}, one has
 \begin{equation}
\langle\overline{\delta^1(\Delta)}\rangle=\frac{F(\alpha-1)}{(2-\alpha)(3-\alpha)} T^{2-\alpha}\Delta.
\end{equation}
It is obvious that the generalized Einstein relation does not hold as Eq. \eqref{AER2} shows for $1<\alpha<2$.

For the second moment of the aging stochastic process, one needs to know the velocity correlation function $\langle v(t_1)v(t_2)\rangle$. From the expression of the velocity process $v(t)$ in Eq. \eqref{LWt}, we find the velocity correlation function contains two terms
\begin{equation}\label{TwoParts}
\begin{split}
\langle v(t_1)v(t_2)\rangle=\langle v(t_1)v(t_2)\rangle_1+\langle v(t_1)v(t_2)\rangle_2,
\end{split}
\end{equation}
where the first term comes from the constant force $F$, the second term is contributed by the impact of noise $\xi(s)$, being the same as the one of free L\'{e}vy walk \cite{FroembergBarkai:2013,WangChenDeng:2019}, and the cross term vanishes due to the zero mean of Gaussian noise $\xi(s)$. Considering the complexity of the explicit velocity correlation function in Eq. \eqref{TwoParts}, we present it in scaling form
\begin{equation}\label{01}
\begin{split}
\langle v(t)v(t+\tau)\rangle&\simeq C_1t^{\nu_1-2}\phi_1\left(\frac{\tau}{t}\right)+C_2t^{\nu_2-2}\phi_2\left(\frac{\tau}{t}\right),
\end{split}
\end{equation}
and resort to the generalized Green-Kubo formula on the two terms, respectively.

\subsection{$0<\alpha<1$}

The form of the two scaling functions and the corresponding parameters in Eq. \eqref{01} have been obtained in Ref. \cite{ChenWangDeng:2019-3}.
For $0<\alpha<1$, both the two scaling functions tend to constant when $q\rightarrow0$, i.e.,
\begin{equation}
\phi_1(q) \simeq c_1=\Gamma(\alpha)\Gamma(3-\alpha)/2,
\end{equation}
and
\begin{equation}\label{phi}
\phi_2(q)\simeq c_2=\Gamma(\alpha)\Gamma(1-\alpha).
\end{equation}
Other parameters are $\nu_1=4$, $\nu_2=2$, $C_1=\frac{F^2}{\Gamma(1-\alpha)\Gamma(\alpha)}$, and $C_2=\frac{D}{\gamma}\frac{1}{\Gamma(1-\alpha)\Gamma(\alpha)}$.
In addition, corresponding to the velocity correlation function $\langle v(t_1)v(t_2)\rangle$ in Eq. \eqref{TwoParts}, the variance of velocity process also consists of two terms, the asymptotic expressions of which are $\langle v^2(t)\rangle_{i}\propto t^{\beta_{i}}$ with $\beta_1=2$ and $\beta_2=0$.

By virtue of the generalized Green-Kubo formula as well as the  Eqs. \eqref{4} and \eqref{5}, one arrives at the second moment of the aging L\'{e}vy walk
\begin{equation}
\begin{split}
\langle x_{t_a}^2(t)\rangle\simeq2 D_{\nu_1}^{t/t_a}t^{\nu_1}+2 D_{\nu_2}^{t/t_a}t^{\nu_2}
\end{split}
\end{equation}
with the diffusion coefficients
\begin{equation}
\begin{split}
D_{\nu_1}^{t/t_a}=\left\{
    \begin{array}{ll}
    F^2\left[\frac{\alpha(1-\alpha)(2-\alpha)(3+\alpha)}{144}+\frac{(1-\alpha)^2}{8}\right], & t\gg t_a, \\[4pt]
    \frac{F^2(1-\alpha)(2-\alpha)}{4}\left(\frac{t_a}{t}\right)^2, & t\ll t_a,
\end{array}
  \right.
\end{split}
\end{equation}
and
\begin{equation}
\begin{split}
D_{\nu_2}^{t/t_a}=\left\{
    \begin{array}{ll}
    \frac{D}{2\gamma}(1-\alpha), & t\gg t_a, \\[4pt]
    \frac{D}{2\gamma}, & t\ll t_a.
\end{array}
  \right.
\end{split}
\end{equation}
Then subtracting the square of the first moment in Eq. \eqref{mean1}, we obtain the EAMSD for weak aging case with $t\gg t_a$,
\begin{equation}\label{LWEAMSD1-1}
\begin{split}
\langle \Delta x^2_{t_a}(t)\rangle&\simeq \frac{F^2\alpha(1-\alpha)(2-\alpha)(3+\alpha)}{72}t^4\\
&~~~~~+\frac{D}{\gamma}(1-\alpha)t^2,
\end{split}
\end{equation}
being the same as the non-aging case \cite{ChenWangDeng:2019-3}. The first term $t^4$, coming from the constant force $F$, dominates the diffusion behavior. We maintain the sub-leading term $t^2$ to show the particle's intrinsic diffusion [i.e., free aging L\'{e}vy walk in Eq. \eqref{freeEAMSD1}]. Therefore, the constant force $F$ enhances the diffusion behavior of L\'{e}vy walk from $t^2$ to $t^4$ in weak aging case.

On the other hand, for strong aging case with $t\ll t_a$, it holds that
\begin{equation}\label{ALWEAMSD1}
\begin{split}
&\langle \Delta x^2_{t_a}(t)\rangle\\&\simeq
\frac{F^2(1-\alpha)(2-\alpha)}{2}t_a^2t^2+\frac{D}{\gamma}t^2-F^2(1-\alpha)^2t_a^2t^2\\
&\simeq\frac{F^2\alpha(1-\alpha)}{2}t_a^2t^2,
\end{split}
\end{equation}
which presents the ballistic diffusion behavior $\propto t^2$ as the free aging L\'{e}vy walk in Eq. \eqref{freeEAMSD1}. Only the diffusion coefficient is increased by the aging time $t_a$. The second line of Eq. \eqref{ALWEAMSD1} implies that the constant force $F$ is the key of resulting in an aging phenomenon, since $F=0$ yields the result $D t^2/\gamma$, which is independent of the aging time $t_a$.

Based on the definition of TAMSD in Eq. \eqref{TAdefination}, the ensemble-averaged TAMSD of the aging L\'{e}vy walk in the constant force field described by Eq. \eqref{LW_force} contains three parts
\begin{equation}
\begin{split}
\langle \overline{\delta_{t_a}^2(\Delta)}\rangle&=\frac{1}{T-\Delta}\int_{t_a}^{t_a+T-\Delta} \langle (x(t+\Delta)-x(t))^2\rangle dt \\
&~~~~-\frac{1}{T-\Delta}\int_{t_a}^{t_a+T-\Delta} \langle x(t+\Delta)-x(t)\rangle^2dt\\
&=\langle \overline{\delta_{t_a}^2(\Delta)}\rangle_1+\langle \overline{\delta_{t_a}^2(\Delta)}\rangle_2-\langle \overline{\delta_{t_a}^2(\Delta)}\rangle_3,
\end{split}
\end{equation}
where the first two terms ($i=1,2$)
\begin{equation}
\begin{split}
\langle \overline{\delta_{t_a}^2(\Delta)}\rangle_{i}=\frac{1}{T-\Delta}\int_{t_a}^{t_a+T-\Delta} \langle (x(t+\Delta)-x(t))^2\rangle_{i}dt
\end{split}
\end{equation}
come from the two parts of velocity correlation function $\langle v(t_1)v(t_2)\rangle_i$, respectively, and the third term is
\begin{equation}\label{3}
\begin{split}
\langle \overline{\delta_{t_a}^2(\Delta)}\rangle_3=\frac{1}{T-\Delta}\int_{t_a}^{t_a+T-\Delta} \langle x(t+\Delta)-x(t)\rangle^2dt.
\end{split}
\end{equation}
By use of the general expression in Eq. \eqref{TA} and the corresponding parameters below Eq. \eqref{01}, one obtains the sum of the first two parts of the ensemble-averaged TAMSD for $0<\alpha<1$
\begin{equation}\label{ALWTAMSD-Add1}
\begin{split}
\sum_{i=1}^2 \langle \overline{\delta_{t_a}^2(\Delta)}\rangle_i=\left\{
    \begin{array}{ll}
    \frac{F^2(1-\alpha)(2-\alpha)}{6}T^2\Delta^2+\frac{D}{\gamma}\Delta^2, & T\gg t_a, \\[4pt]
    \frac{F^2(1-\alpha)(2-\alpha)}{2}t_a^2\Delta^2+\frac{D}{\gamma}\Delta^2, & T\ll t_a.
\end{array}
  \right.
\end{split}
\end{equation}
The third part is obtained by use of the first moment in Eq. \eqref{mean1}
\begin{equation}
\begin{split}
\langle \overline{\delta_{t_a}^2(\Delta)}\rangle_3\simeq\left\{
    \begin{array}{ll}
    \frac{F^2(1-\alpha)^2}{3}T^2\Delta^2, & T\gg t_a, \\[4pt]
    F^2(1-\alpha)^2t_a^2\Delta^2, & T\ll t_a.
\end{array}
  \right.
\end{split}
\end{equation}
Therefore, the ensemble-averaged TAMSD is
\begin{equation}\label{ALWTAMSD1}
\begin{split}
\langle \overline{\delta_{t_a}^2(\Delta)}\rangle\simeq\left\{
    \begin{array}{ll}
    \frac{F^2\alpha(1-\alpha)}{6}T^2\Delta^2, & T\gg t_a, \\[4pt]
    \frac{F^2\alpha(1-\alpha)}{2}t_a^2\Delta^2, & T\ll t_a,
\end{array}
  \right.
\end{split}
\end{equation}
both showing the ballistic diffusion behavior $\propto \Delta^2$ as the free aging L\'{e}vy walk in Eq. \eqref{freeTAMSD1} does. The difference between weak and strong aging cases are embodied by the $T$- or $t_a$-dependent diffusion coefficient. With the increase of measurement time $T$ or aging time $t_a$, the ensemble-averaged TAMSD becomes larger. Similar to the result of EAMSD in Eq. \eqref{ALWEAMSD1}, the constant force $F$ is the key of resulting in the aging phenomenon. Otherwise, from Eq. \eqref{ALWTAMSD-Add1} or Eq. \eqref{freeTAMSD1}, the force-free case with $F=0$ yields the result $D \Delta^2/\gamma$ independent of aging time $t_a$.

\subsection{$1<\alpha<2$}

Let us turn to the sub-ballistic superdiffusive L\'{e}vy walk with $1<\alpha<2$. The scaling functions $\phi_i(q)$ yield different asymptotic behaviors from the case with $0<\alpha<1$. In detail, it holds that \cite{ChenWangDeng:2019-3}
\begin{equation}
\phi_1(q)\simeq \frac{2-\alpha}{3-\alpha},\quad \phi_2(q)\simeq q^{1-\alpha},
\end{equation}
when $q\rightarrow0$. Other parameters are $\nu_1=5-\alpha$, $\nu_2=3-\alpha$, $C_1=\frac{F^2(\alpha-1)}{2-\alpha}$, $C_2=\frac{D}{\gamma}$, $c_1=\frac{2-\alpha}{3-\alpha}$ and $c_2=1$. The two parts of the variance of velocity process are $\langle v^2(t)\rangle_{i}\propto t^{\beta_{i}}$ with $\beta_1=3-\alpha$ and $\beta_2=0$.
Substituting these parameters into the generalized Green-Kubo formula, we obtain the asymptotic expression of the aging EAMSD, which is, for the weak aging case with $t\gg t_a$,
\begin{equation}\label{A2EAMSD1}
\begin{split}
&\langle \Delta x^2_{t_a}(t)\rangle\\
&\simeq
    \frac{F^2(\alpha-1)}{(4-\alpha)(5-\alpha)}t^{5-\alpha}
    +\frac{2D(\alpha-1)}{\gamma(2-\alpha)(3-\alpha)}t^{3-\alpha},
\end{split}
\end{equation}
and for the strong aging case with $t\ll t_a$,
\begin{equation}\label{A2EAMSD2}
\begin{split}
&\langle \Delta x^2_{t_a}(t)\rangle\\
&\simeq
  \frac{F^2(\alpha-1)}{3-\alpha}t_a^{3-\alpha}t^2
 +\frac{2D}{\gamma(2-\alpha)(3-\alpha)}t^{3-\alpha}.
\end{split}
\end{equation}
In the two equations above, the first terms containing the constant force $F$ are also the dominating ones, which implies the constant force enhances the diffusion behavior of the aging process.
The second terms containing diffusivity $D$ come from the particle's intrinsic motion. We maintain the sub-leading terms to show the results of the force-free case as Eq. \eqref{freeEAMSD2} shows and to reveal the effects of the constant force through a direct comparison. For $F=0$, both weak and strong aging cases exhibit the sub-ballistic superdiffusion $t^{3-\alpha}$. But in the effect of the constant force, the diffusion behavior is enhanced in different way for weak and strong aging cases, being $t^{5-\alpha}$ and $t_a^{3-\alpha}t^2$, respectively.

Similarly, the ensemble-averaged TAMSD is, for weak aging case with $T\gg t_a$,
\begin{equation}\label{A2TAMSD1}
\begin{split}
&\langle \overline{\delta_{t_a}^2(\Delta)}\rangle\\
&\simeq \frac{F^2(\alpha-1)}{(3-\alpha)(4-\alpha)}T^{3-\alpha}\Delta^2
+\frac{2D}{\gamma(2-\alpha)(3-\alpha)}\Delta^{3-\alpha},
    \end{split}
\end{equation}
and for strong aging case with $T\ll t_a$,
\begin{equation}\label{A2TAMSD2}
\begin{split}
&\langle \overline{\delta_{t_a}^2(\Delta)}\rangle\\
&\simeq \frac{F^2(\alpha-1)}{3-\alpha}t_a^{3-\alpha}\Delta^2
+\frac{2D}{\gamma(2-\alpha)(3-\alpha)}\Delta^{3-\alpha}.
    \end{split}
\end{equation}
It can also be found that the first terms in Eqs. \eqref{A2TAMSD1} and \eqref{A2TAMSD2}, contributed by the constant force, play the leading role due to $\Delta\ll T$. Therefore, the constant force enhances the diffusion behavior with respect to the TAMSD from $\Delta^{3-\alpha}$ to $\Delta^2$, being the same as Eq. \eqref{ALWTAMSD1} for $0<\alpha<1$. The difference is embodied by the $T$- or $t_a$- dependent diffusivity, which also implies the aging phenomenon of TAMSD.
All the MSDs in Eqs. \eqref{A2EAMSD1}-\eqref{A2TAMSD2} can recover to the case of the free aging L\'{e}vy walk in Sec. \ref{two} by taking $F=0$.

\begin{figure}
  \centering
  \includegraphics[scale=0.2]{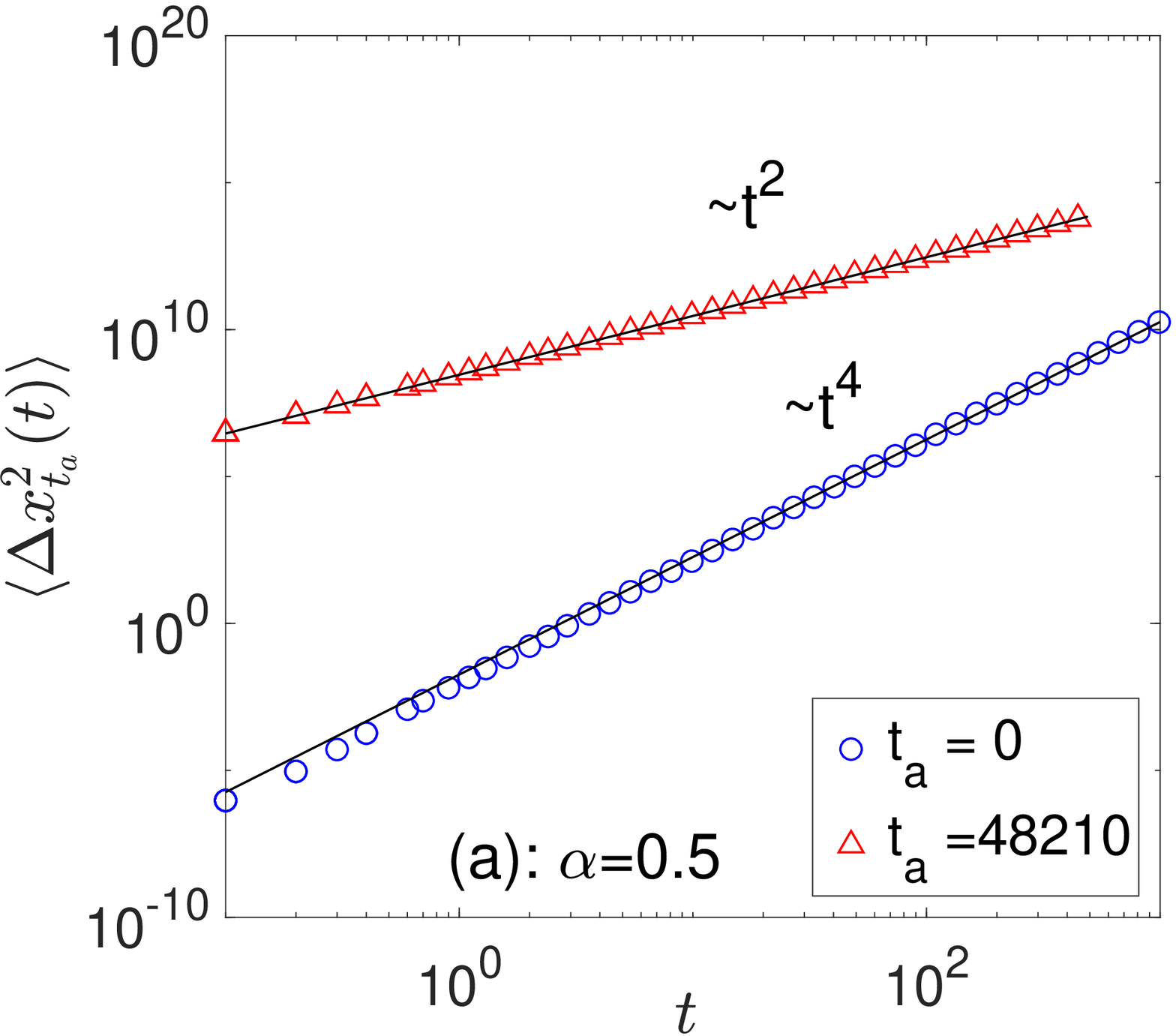}
  \includegraphics[scale=0.2]{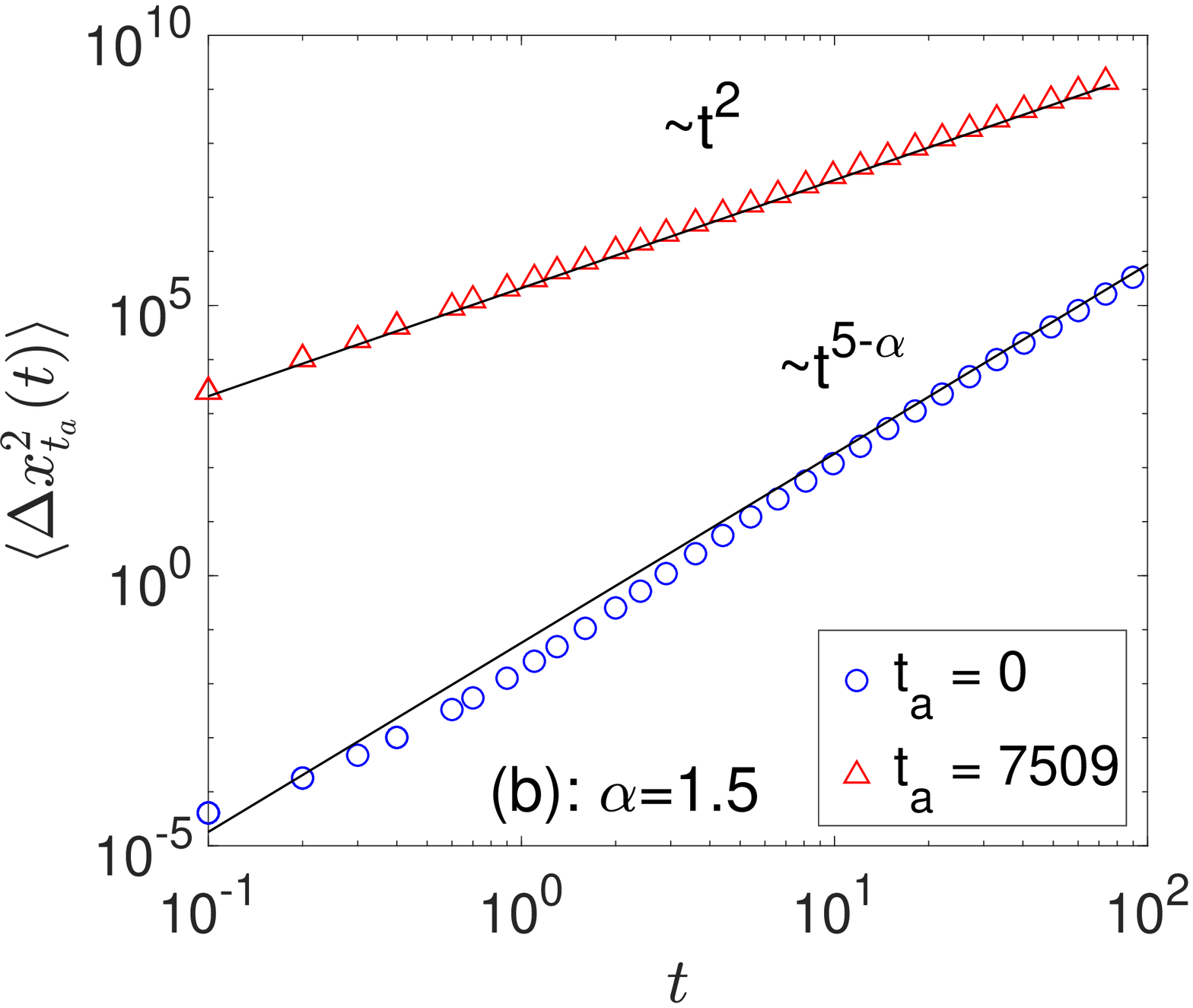}\\
  \includegraphics[scale=0.2]{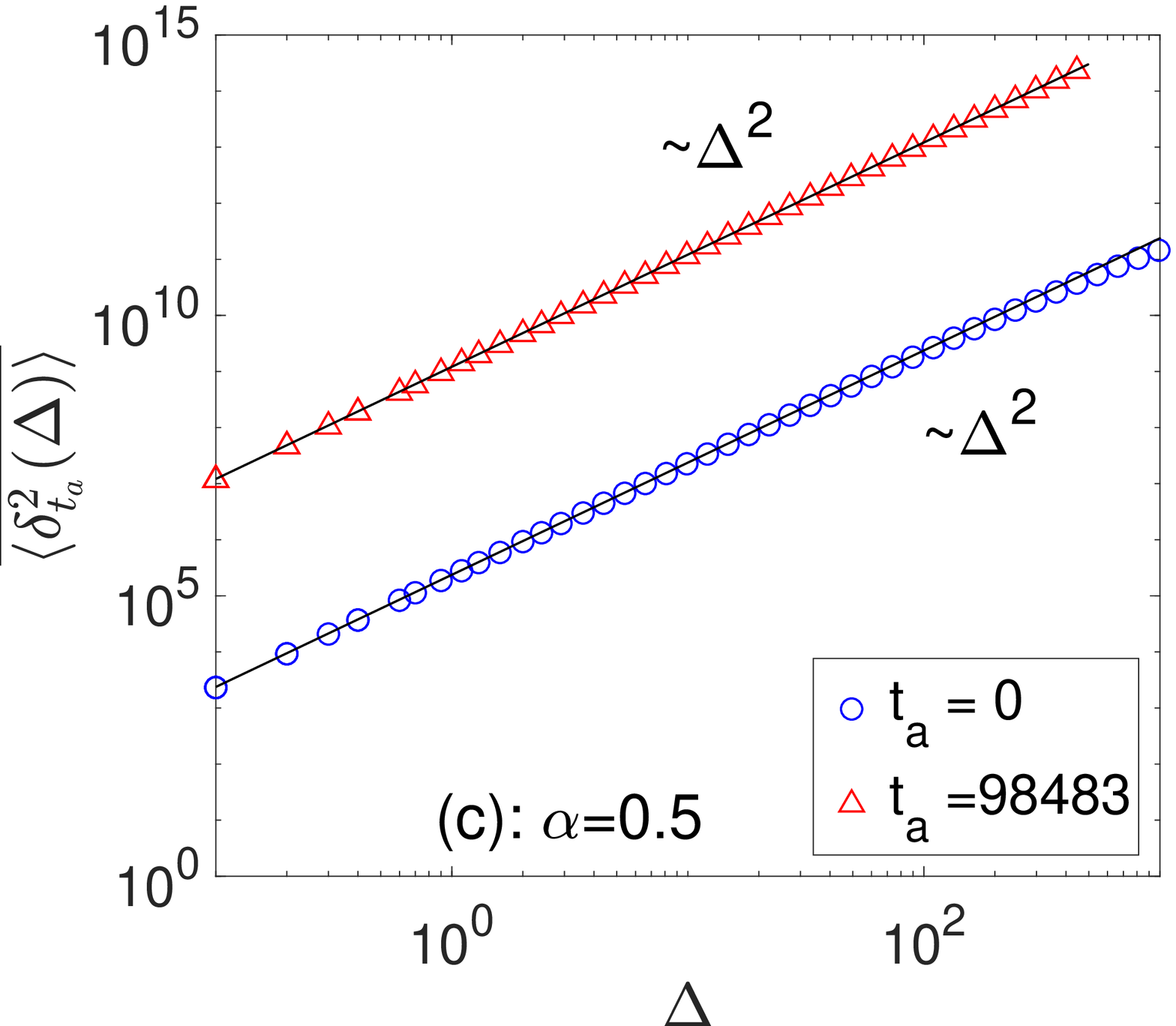}
  \includegraphics[scale=0.2]{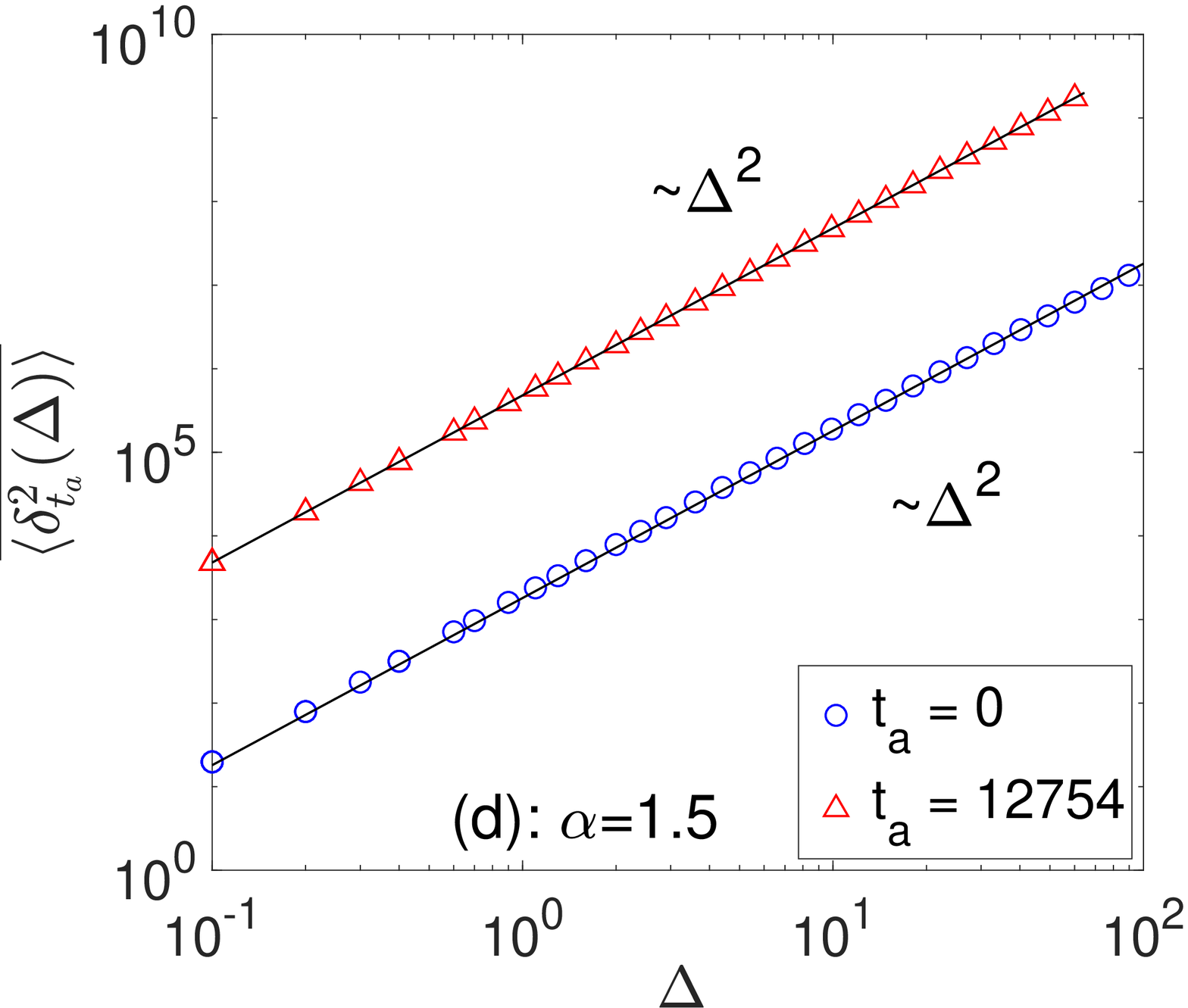}
  \caption{EAMSDs and ensemble-averaged TAMSDs of the aging L\'{e}vy walk under the effect of the constant force $F$ for different $\alpha$. Blue circles and red triangles represent the simulation results of weak aging and strong aging cases, respectively. The black solid lines are the theoretical results with asymptotic forms in Eqs. \eqref{LWEAMSD1-1}, \eqref{ALWEAMSD1} and \eqref{ALWTAMSD1} for $0<\alpha<1$, and Eqs. \eqref{A2EAMSD1}-\eqref{A2TAMSD2} for $1<\alpha<2$. Other parameters: $F=1$, $D=1$, $\gamma=2$.
  The upper panels are for EAMSDs while the lower panels are for ensemble-averaged TAMSDs with $\alpha=0.5$ and $\alpha=1.5$. The simulation results agree with the theoretical ones for all cases.}\label{fig1}
\end{figure}

The simulations of the EAMSDs and ensemble-averaged TAMSDs are presented in Fig. \ref{fig1}, where we put the weak aging (blue circles) and strong aging (red triangles) cases in one panel for comparison. For the EAMSD, the age $t_a$ suppresses the diffusion behavior from $t^4$ to $t^2$ for $0<\alpha<1$ and from $t^{5-\alpha}$ to $t^2$ for $1<\alpha<2$, but increases the diffusion coefficients. While for the ensemble-averaged TAMSD, the ballistic diffusion behaviors are observed for both weak and strong aging cases. The age time $t_a$ only increases the diffusion coefficients for both cases with $0<\alpha<1$ and $1<\alpha<2$.

From the aspect of ergodic properties of the aging L\'{e}vy walk under a constant force $F$, based on Eqs. \eqref{LWEAMSD1-1}, \eqref{ALWEAMSD1}, \eqref{ALWTAMSD1} for $0<\alpha<1$, and Eqs. \eqref{A2EAMSD1}-\eqref{A2TAMSD2} for $1<\alpha<2$, we find the difference between the EAMSD and the ensemble-averaged TAMSD for weak aging cases. But they are consistent for strong aging cases, i.e.,
\begin{equation}\label{EATAEQ}
   \langle \Delta x_{t_a}^2(\Delta)\rangle \simeq \langle\overline{\delta_{t_a}^2(\Delta)}\rangle
\end{equation}
 for $t_a\gg T\gg \Delta$, which shows that the strong aging seemingly yields an ergodic phenomenon. In fact, the TAMSD should converges to a deterministic constant for large measurement time $T$ for an ergodic system. For $1<\alpha<2$, the mean sojourn time of L\'{e}vy walk is finite, and the individual trajectories become self-averaging at sufficiently large time, such that there will be no difference between $\overline{\delta_{t_a}^2(\Delta)}$ obtained from different trajectories and the ensemble-averaged quantity $\langle\overline{\delta_{t_a}^2(\Delta)}\rangle$ \cite{GodecMetzler:2013,FroembergBarkai:2013,WangChenDeng:2019-2}. While for $0<\alpha<1$, the characteristic time scale of L\'{e}vy walk is infinite, then the individual TAMSD $\overline{\delta_{t_a}^2(\Delta)}$ is irreproducible and inequivalent with the corresponding EAMSD, which implies the ergodicity breaking in this case.

\section{Aging L\'{e}vy walk in time-dependent periodic force field}\label{four}

The L\'{e}vy-walk-like Langevin dynamics in a general time-dependent force field $F(t)$ is expressed as \cite{ChenDeng:2021}
\begin{equation}\label{LW_tforce}
\begin{split}
    \frac{d}{d t}x(t)&=v(t),\\
    \frac{d}{d s}v(s)&=-\gamma v(s) +F(t(s)) \eta(s)+\xi(s),\\
    \frac{d}{d s}t(s)&= \eta(s).
\end{split}
\end{equation}
Similar to Eq. \eqref{LW_force}, the force term $F(t(s(t)))=F(t)$ multiplied by the L\'{e}vy noise $\eta(s)$ in the second sub-equation, implies the time-dependent force acts on the system throughout all physical time $t$, rather than the operational time $s$ \cite{MagdziarzWeronKlafter:2008}.
Replacing the operational time $s$ by the inverse subordinator $s(t)$, the velocity process can be solved and expressed in physical time:
\begin{equation}
\begin{split}
v(t)&=\int_0^t e^{-\gamma(s(t)-s(t'))}F(t')dt'  \\
&~~~+\int_0^t e^{-\gamma(s(t)-s(t'))}\xi(s(t'))ds(t'),
\end{split}
\end{equation}
which presents a similar form as Eq. \eqref{LWt} shows. The first term comes from the external time-dependent force, and the second term from the random force $\xi(s)$, which corresponds to the free L\'{e}vy walk.

In the following discussions, we choose the the time-dependent periodic force $F(t)=f_0 \sin(\omega t)$.
The first moment of velocity process for large time $t$ is \cite{ChenDeng:2021}
\begin{equation}
\begin{split}
\langle v(t)\rangle  \simeq \left\{
    \begin{array}{ll}
    \frac{f_0}{\omega \gamma\Gamma(1-\alpha)}t^{-\alpha}, &~~ 0<\alpha<1, \\[4pt]
    \frac{f_0}{\omega\gamma}t^{-\alpha}, &~~ 1<\alpha<2,
\end{array}
  \right.
\end{split}
\end{equation}
which are decaying at the rate $t^{-\alpha}$ for different $\alpha$. Therefore, for weak and strong aging cases, the aging first moment is
\begin{equation}\label{mean-x2}
\begin{split}
\langle x_{t_a}(t)\rangle  \propto \left\{
    \begin{array}{ll}
    t^{1-\alpha}, &t\gg t_a, \\[4pt]
    t_a^{-\alpha}t, &t\ll t_a.
\end{array}
  \right.
\end{split}
\end{equation}
Similar to the first moment of velocity process, the coefficients have different expressions for $0<\alpha<1$ and $1<\alpha<2$. We omit them here since the first moments can be neglected when calculate the EAMSD later.

From Ref. \cite{ChenDeng:2021}, we know the time-dependent periodic force $F(t)$ slightly enhances the diffusion behavior through an additional diffusivity, which can be found from the same enhancement in the velocity correlation function of the process $v(t)$ described by Eq. \eqref{LW_tforce}: \cite{ChenDeng:2021}
\begin{equation}\label{VCF2}
\begin{split}
&\langle v(t_1)v(t_2)\rangle \\
&\simeq \left\{
\begin{array}{ll}
  \left(D_1+\frac{D}{\gamma}\right)\frac{\sin(\pi\alpha)}{\pi}B\left(\frac{t_1}{t_2};\alpha,1-\alpha \right),  & ~0<\alpha<1,  \\[5pt]
  \left(D_2+\frac{D}{\gamma}\right)\left((t_2-t_1)^{1-\alpha}-t_2^{1-\alpha}\right), &~ 1<\alpha<2,
\end{array}\right.
\end{split}
\end{equation}
where $t_1$, $t_2$ are large and $t_1<t_2$. The expressions of coefficients $D_1$ and $D_2$ are complex and we put them in Appendix \ref{App1}.
Similar to the constant force case, we firstly transform the velocity correlation function above into the scaling form as Eq. \eqref{vv} shows:
\begin{equation}
\begin{split}
\langle v(t)v(t+\tau)\rangle
\simeq \left\{
\begin{array}{ll}
  C_3t^{\nu_3-2}\phi_3\left(\frac{\tau}{t}\right),  & ~0<\alpha<1,  \\[5pt]
  C_4t^{\nu_4-2}\phi_4\left(\frac{\tau}{t}\right), &~ 1<\alpha<2,
\end{array}\right.
\end{split}
\end{equation}
where the scaling functions are
\begin{equation}
  \begin{split}
    &\phi_3(q)=B\left(\frac{1}{1+q};\alpha,1-\alpha \right)\simeq B(\alpha,1-\alpha), \\
    &~~~~~~\phi_4(q)=q^{1-\alpha}-(1+q)^{1-\alpha}\simeq q^{1-\alpha},
  \end{split}
\end{equation}
as $q\rightarrow0$, and other parameters are $\nu_3=2$, $\nu_4=3-\alpha$, $C_3=\left(D_1+\frac{D}{\gamma}\right)\frac{\sin(\pi\alpha)}{\pi}$,~
~$C_4=\left(D_2+\frac{D}{\gamma}\right)$, $c_3=B(\alpha,1-\alpha)$,~$c_4=1$, $\beta_3=\beta_4=0$. Then based on the generalized Green-Kubo formula, we obtain the aging EAMSD for $0<\alpha<1$
\begin{equation}\label{A3EAMSD1}
\langle \Delta x^2_{t_a}(t)\rangle\simeq \left\{
\begin{array}{ll}
  \left(D_1+\frac{D}{\gamma}\right)(1-\alpha)t^2,  & t\gg t_a, \\[5pt]
  \left(D_1+\frac{D}{\gamma}\right)t^2, & t\ll t_a,
\end{array}\right.
\end{equation}
and for $1<\alpha<2$
\begin{equation}\label{EAtime}
\langle \Delta x^2_{t_a}(t)\rangle\simeq \left\{
\begin{array}{ll}
\left(D_2+\frac{D}{\gamma}\right)
\frac{2(\alpha-1)}{(2-\alpha)(3-\alpha)}t^{3-\alpha},  & t\gg t_a, \\[5pt]
\left(D_2+\frac{D}{\gamma}\right)
\frac{2}{(2-\alpha)(3-\alpha)}t^{3-\alpha}, & t\ll t_a.
\end{array}\right.
\end{equation}
Note that the square of the first moment of displacement $\langle x_{t_a}(t)\rangle^2$ in Eq. \eqref{mean-x2}, which is far less than the results in Eqs. \eqref{A3EAMSD1} and \eqref{EAtime}, has been omitted.
Although the specific expressions of the aging EAMSDs are independent of the aging time $t_a$, and are only different from the non-aging case in Eqs. \eqref{freeEAMSD1} and \eqref{freeEAMSD2} by the additional coefficients $D_1$ and $D_2$, respectively, the EAMSDs are different for weak and strong aging cases.

\begin{figure}
  \centering
  \includegraphics[scale=0.2]{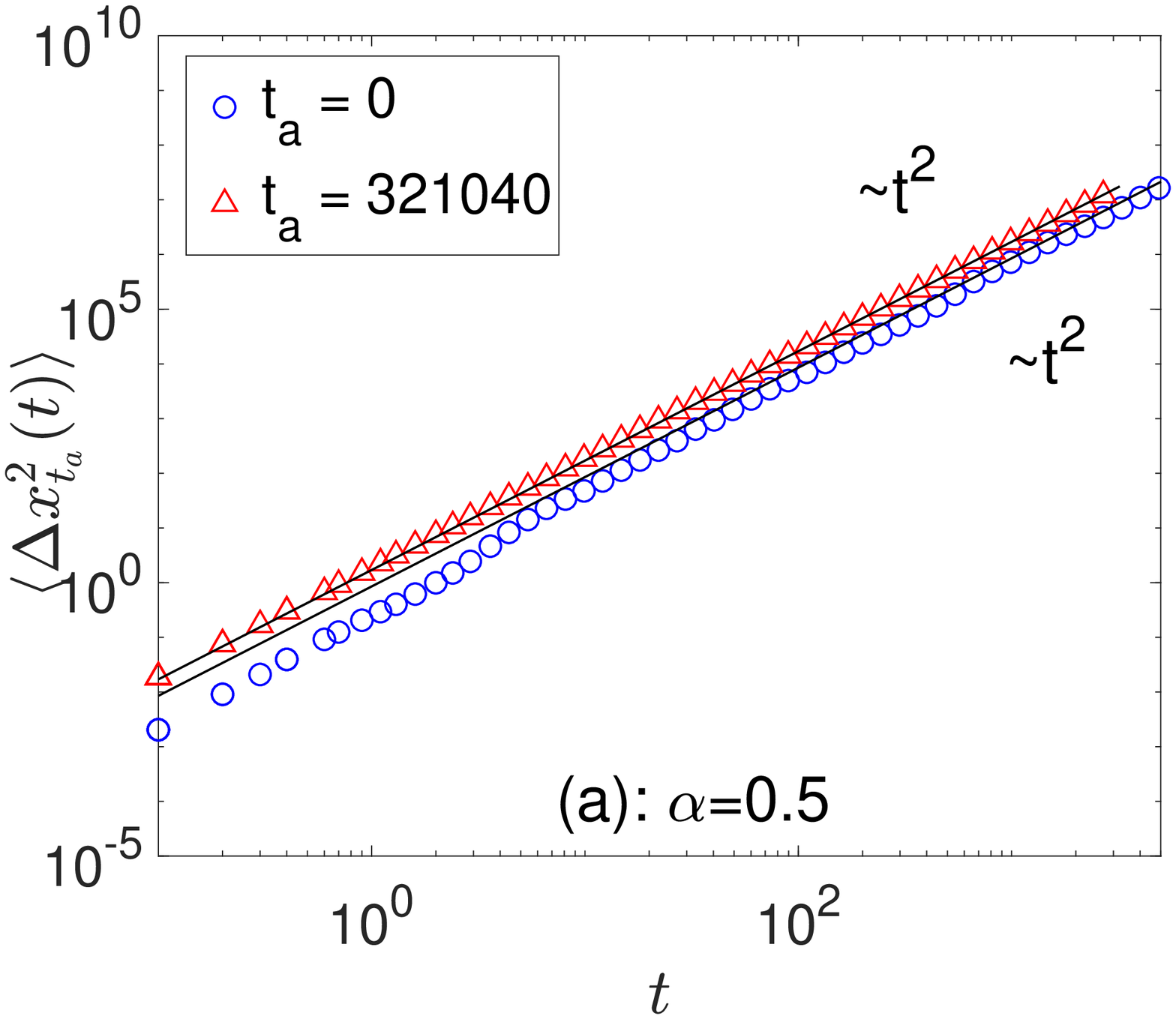}
  \includegraphics[scale=0.2]{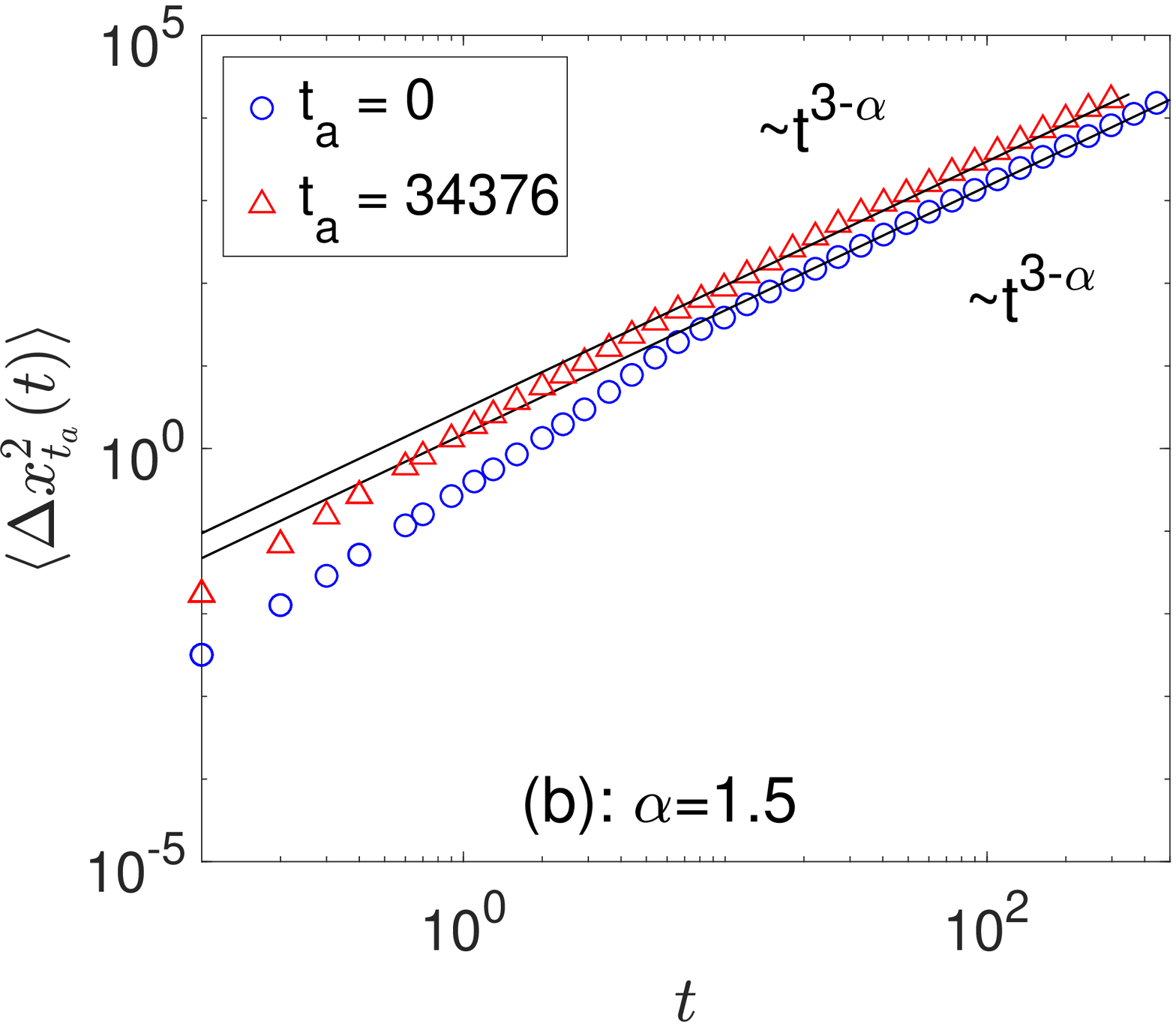}\\
  \includegraphics[scale=0.2]{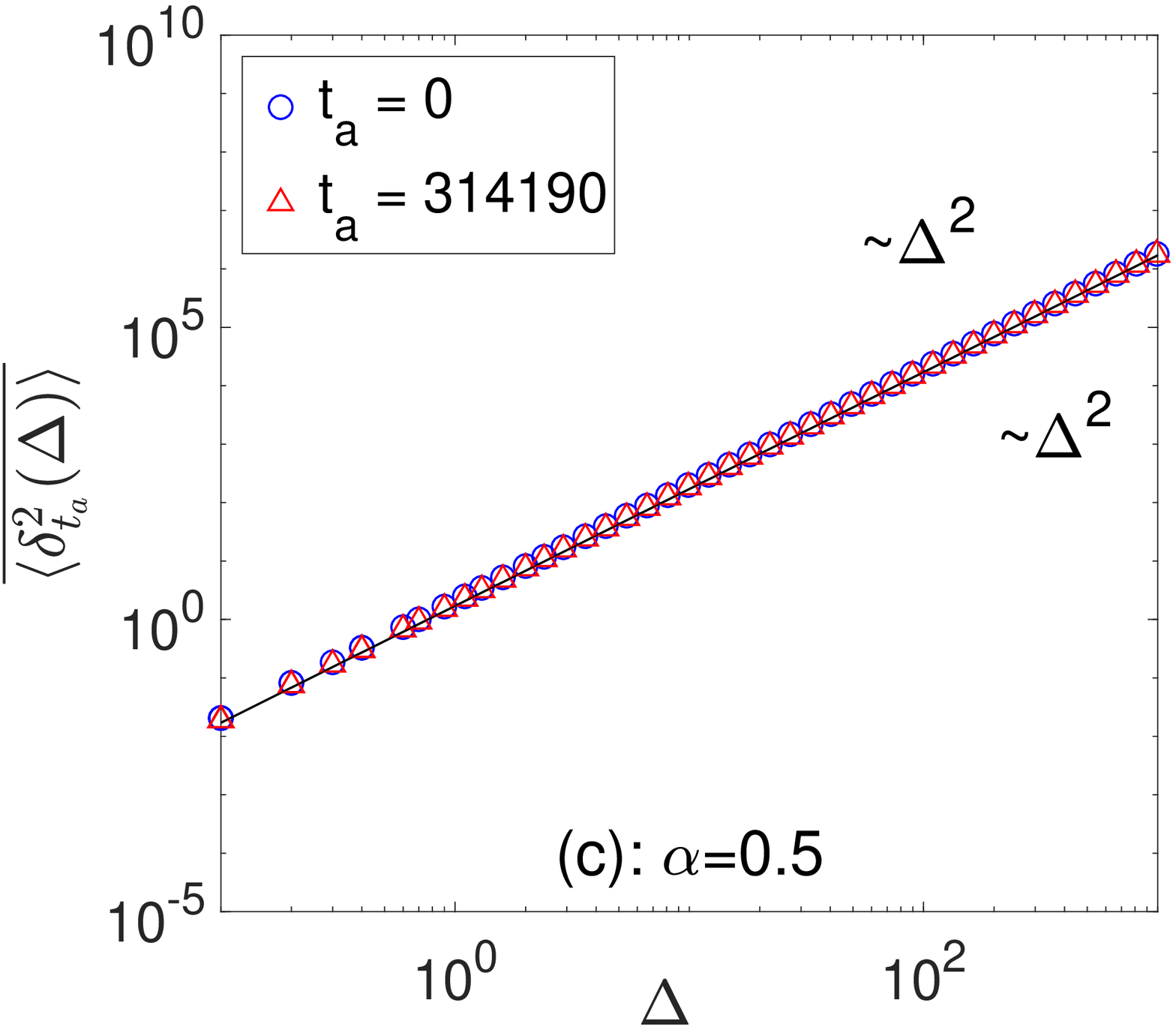}
  \includegraphics[scale=0.2]{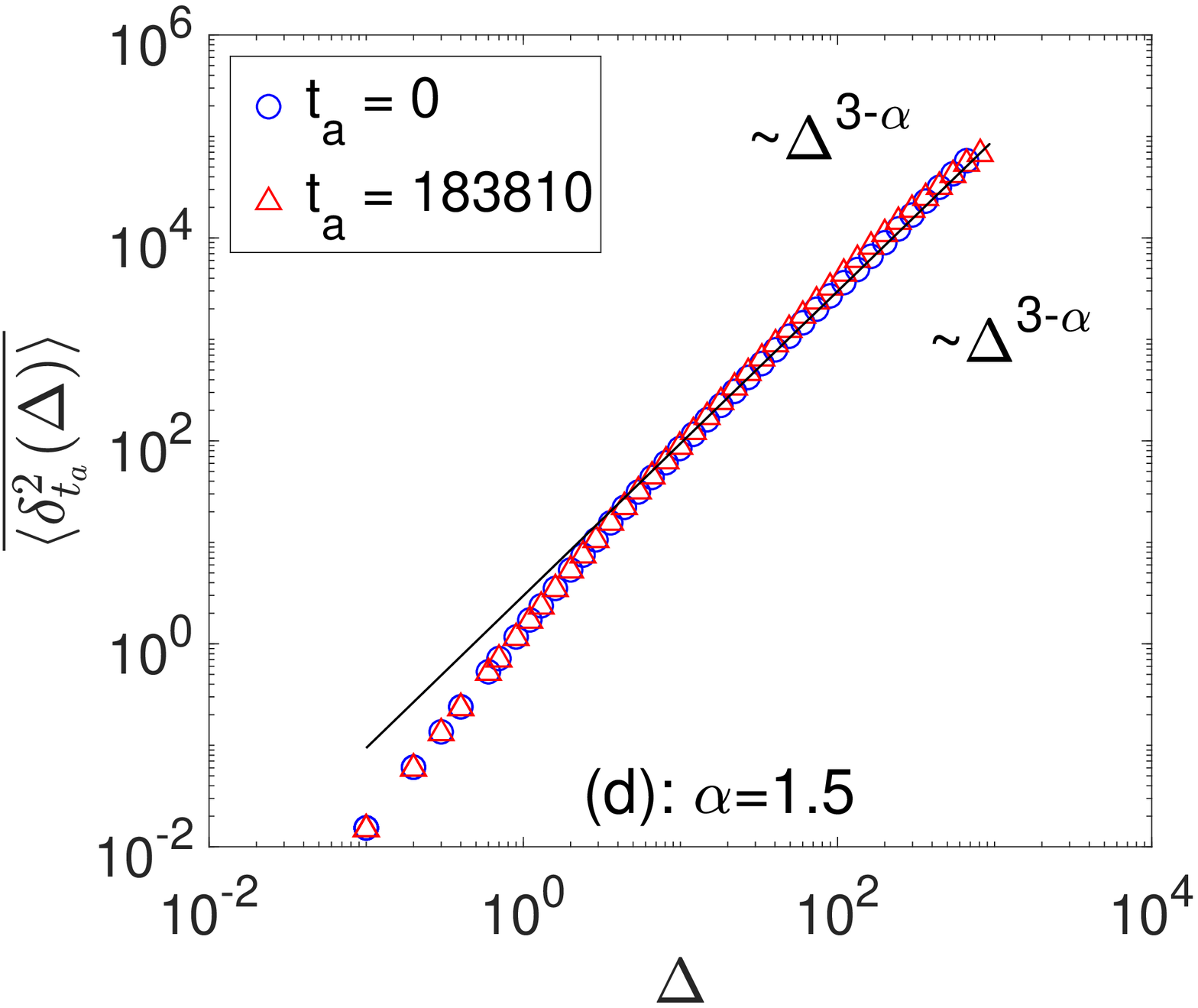}
  \caption{EAMSDs and ensemble-averaged TAMSDs of the aging L\'{e}vy walk under the effect of the time-dependent periodic force $F(t)=f_0\sin(\omega t)$ for different $\alpha$. Blue circles and red triangles represent the simulation results of weak aging and strong aging cases, respectively. The black solid lines are the theoretical results with asymptotic forms in Eqs. \eqref{A3EAMSD1}, \eqref{TAMSD3-1} and \eqref{TAMSD3-1} for $0<\alpha<1$, and Eqs. \eqref{EAtime} and \eqref{a} for $1<\alpha<2$. Other parameters: $f_0=1$, $D=1$, $\gamma=2$, $\omega=0.5$.
  The upper panels are for EAMSDs while the lower panels are for ensemble-averaged TAMSDs with $\alpha=0.5$ and $\alpha=1.5$. The simulation results agree with the theoretical ones for all cases.}\label{fig2}
\end{figure}

On the other hand, for the case with $0<\alpha<1$, based on the generalized Green-Kubo formula, the ensemble-averaged TAMSD is
\begin{equation}\label{TAMSD3-1}
\begin{split}
&\langle \overline{\delta_{t_a}^2(\Delta)}\rangle\\
&\simeq \left(D_1+\frac{D}{\gamma}\right)\Delta^2
-\frac{f_0^2}{\omega^2\gamma^2\Gamma^2(1-\alpha)(1-2\alpha)}T^{-2\alpha}\Delta^2\\
&\simeq \left(D_1+\frac{D}{\gamma}\right)\Delta^2
    \end{split}
\end{equation}
for $T\gg t_a$, and
\begin{equation}\label{TAMSD3-2}
\begin{split}
&\langle \overline{\delta_{t_a}^2(\Delta)}\rangle\\
&\simeq \left(D_1+\frac{D}{\gamma}\right)\Delta^2
-\frac{f_0^2}{\omega^2\gamma^2\Gamma^2(1-\alpha)}t_a^{-2\alpha}\Delta^2\\
&\simeq \left(D_1+\frac{D}{\gamma}\right)\Delta^2
    \end{split}
\end{equation}
for $T\ll t_a$. Both the second terms in the second lines of Eqs. \eqref{TAMSD3-1} and \eqref{TAMSD3-2} are contributed by the square of the aging first moment. Although they are different for weak and strong cases,
they are far less than the dominating term $\Delta^2$ and can be omitted.
Therefore, we obtain the same asymptotic behavior of the ensemble-averaged TAMSD. Similarly, for $1<\alpha<2$, the aging ensemble-averaged TAMSD is
\begin{equation}\label{a}
\begin{split}
\langle \overline{\delta_{t_a}^2(\Delta)}\rangle
\simeq \left(D_2+\frac{D}{\gamma}\right)
\frac{2}{(2-\alpha)(3-\alpha)}\Delta^{3-\alpha}
    \end{split}
\end{equation}
for both $T\gg t_a$ and $T\ll t_a$. Similar to the force-free case of the aging L\'{e}vy walk in Eqs. \eqref{freeTAMSD1} and \eqref{freeTAMSD2}, the ensemble-averaged TAMSDs present the same asymptotic behaviors for both weak and strong cases in the presence of the time-dependent periodic force.

For both EAMSD and ensemble-averaged TAMSD, the time-dependent periodic force $F(t)$ enhances the diffusion  by adding an additional diffusivity without changing the diffusion behaviors, which is the main difference from the constant force case. The latter strengthes the diffusion behaviors and yields the explicit dependence on age time $t_a$ with respect to both EAMSD and ensemble-averaged TAMSD in the strong aging cases in Sec. \ref{three}. For both cases with constant force and time-dependent periodic force, however, the EAMSD and the ensemble-averaged TAMSD have the same asymptotic behavior in the strong aging cases as Eq. \eqref{EATAEQ} shows, which can also be observed in other systems \cite{WangChenDeng:2019-2}.

To focus on the aging phenomena of the aging L\'{e}vy walk, we present the simulations of the EAMSDs and ensemble-averaged TAMSDs in Fig. \ref{fig2}, and put the weak aging (blue circles) and strong aging (red triangles) cases in one panel for comparison. With respect to the EAMSD, the age time $t_a$ increases the diffusion coefficient without changing the diffusion behavior for both $0<\alpha<1$ and $1<\alpha<2$. While the ensemble-averaged TAMSDs have the same asymptotic behavior for weak aging and strong aging cases.

\section{Summary}\label{five}

Aging phenomena  have  been found in many kinds of anomalous diffusion processes. In this paper, we focus on the different effects of the external force fields on aging L\'{e}vy walk by considering two typical forces, constant force $F$ and time-dependent periodic force $F(t)=f_0\sin(\omega t)$.
Based on the Langevin equation and the two-point joint PDF of the inverse subordinator, the velocity correlation function, and further the EAMSD and ensemble-averaged TAMSD can be obtained. We find that the constant force is the key of causing the aging phenomena and it enhances the diffusion behavior of the aging L\'{e}vy walk, while the time-dependent periodic force does not.

The main results are summarized in Table \ref{table}. For the aging L\'{e}vy walk in the external constant force field, in the weak aging case $t\gg t_a$, the EAMSD behaves as $t^4$ for $0<\alpha<1$ and $t^{5-\alpha}$ for $1<\alpha<2$, being the same as the non-aging case, which is faster than the force-free case. In the strong aging case $t\ll t_a$, the EAMSD shows the ballistic behavior with respect to the measurement time $t$ as $\propto t_a^2t^2$ for $0<\alpha<1$ and $\propto t_a^{3-\alpha}t^2$ for $1<\alpha<2$. The aging phenomena are embodied by the explicit dependence on the aging time $t_a$.
While for the ensemble-averaged TAMSD, the weak aging case $T\gg t_a$ exhibits $T^2\Delta^2$ for $0<\alpha<1$ and $T^{3-\alpha}\Delta^2$ for $1<\alpha<2$. The former presents the ballistic behavior $\Delta^2$, being
the same as the free aging L\'{e}vy walk. On the contrast, the latter one is also $\Delta^2$, faster than the free aging case $\Delta^{3-\alpha}$. For the strong aging case $T\ll t_a$, the ensemble-averaged TAMSD also shows the ballistic behavior $\Delta^2$, together with the explicit dependence on the aging time $t_a$, implies the aging phenomena.
For the case with the external time-dependent period force field, the EAMSD and ensemble-averaged TAMSD have the same scaling forms for weak aging and strong aging cases, which is similar to the free aging L\'{e}vy walk.

It is very common that the particles are subjected to some kinds of external force fields in the natural world. Many researches focus on the effects of external forces on the subdiffusive CTRW with power-law-distributed waiting times  \cite{MetzlerKlafter:2000,BelBarkai:2005_2,SokolovKlafter:2006,MagdziarzWeronKlafter:2008,EuleFriedrich:2009,AkimotoCherstvyMetzler:2018,ChenWangDeng:2019-2},
where the constant force and time-dependent periodic force present some similarities. More precisely, if these two kinds of forces act on the subdiffusive CTRW throughout both waiting times and jumping moments, then they both behave as decoupled force and do not change the MSDs \cite{ChenWangDeng:2019-2}. This is the generic property of the Galilei invariant diffusion processes \cite{MetzlerKlafter:2000,CairoliKlagesBaule:2018,VotAbadMetzlerYuste:2020}, since the effect of the position-independent force is tantamount to a change of Galilei reference frame. While for superdiffusive L\'{e}vy walk, especially the aging case, the constant force and time-dependent periodic force yield different effects. The results in this paper may bring benefits to the studies of diffusion processes who are affected by different kinds of external forces.

\begin{widetext}

\begin{table}
\centering
\caption{Asymptotic behaviors of EAMSD and ensemble-averaged TAMSD (only maintaining time variables) for aging L\'{e}vy walk in the constant force field and time-dependent period force field.}\label{table}
\begin{tabular}{*{6}{|c}|}
\hline
  Force types & Parameter $\alpha$ & $\langle \Delta x^2_{t_a}(t)\rangle$ & $\langle\overline{\delta_{t_a}^2(\Delta)}\rangle$ & Aging types & Eqs. \\
\hline
  \multirow{4}{*}{Constant force $F$} & \multirow{2}{*}{ $0<\alpha<1$} & $t^4$ & $T^2\Delta^2$ & weak &\eqref{LWEAMSD1-1} and \eqref{ALWTAMSD1}\\
\cline{3-6}
 & & $t_a^2t^2$ & $t_a^2\Delta^2$ & strong&\eqref{ALWEAMSD1} and \eqref{ALWTAMSD1}\\
\cline{2-6}
 & \multirow{2}{*}{ $1<\alpha<2$} & $t^{5-\alpha}$ & $T^{3-\alpha}\Delta^2$ & weak&\eqref{A2EAMSD1} and \eqref{A2TAMSD1}\\
\cline{3-6}
 &  & $t_a^{3-\alpha}t^2$ & $t_a^{3-\alpha}\Delta^2$ &strong&\eqref{A2EAMSD2} and \eqref{A2TAMSD2}\\
\hline
  \multirow{2}{*}{Time-dependent periodic force $F(t)=f_0\sin(\omega t)$} & \multirow{1}{*}{ $0<\alpha<1$} & $t^2$  & $\Delta^2$  &weak and strong&\eqref{A3EAMSD1}, \eqref{TAMSD3-1} and \eqref{TAMSD3-2}\\
\cline{2-6}
 & \multirow{1}{*}{ $1<\alpha<2$} & $t^{3-\alpha}$  & $\Delta^{3-\alpha}$  &weak and strong&\eqref{EAtime} and \eqref{a}\\
\hline
\end{tabular}
\end{table}

\end{widetext}

\begin{acknowledgments}
This work was supported by the National Natural Science
Foundation of China under Grant No. 12105145, the Natural Science Foundation of Jiangsu Province under Grant No. BK20210325.
\end{acknowledgments}

\appendix

\section{Coefficients $D_1$ and $D_2$ in Eq. \eqref{VCF2}}\label{App1}
The explicit expressions of coefficients $D_1$ and $D_2$ are
\begin{equation}
  D_1=\frac{f^2_0\omega^{\alpha-2}(\gamma\cos(\alpha\pi/2)+\omega^\alpha)}{2(\gamma^2+2\gamma\omega^\alpha\cos(\alpha\pi/2)+\omega^{2\alpha})}
\end{equation}
and
\begin{equation}
  D_2=\frac{f^2_0b_1}{2\omega^2(\gamma^2+b_2)},
\end{equation}
where
\begin{equation}
  \begin{split}
    &~~~~~~~~~~~b_1=\omega^2/{(\alpha-1)^2}+|\Gamma(1-\alpha)|^2\omega^{2\alpha} \\
&-|\Gamma(1-\alpha)|\omega^\alpha(\gamma\cos(\alpha\pi/2)+2\omega/(\alpha-1)\sin(\alpha\pi/2)),
  \end{split}
\end{equation}
and
\begin{equation}
  \begin{split}
  &~~~~~~~~~~~b_2=\omega^2/{(\alpha-1)^2}+|\Gamma(1-\alpha)|^2\omega^{2\alpha} \\
&-2|\Gamma(1-\alpha)|\omega^\alpha(\gamma\cos(\alpha\pi/2)+\omega/(\alpha-1)\sin(\alpha\pi/2)).
  \end{split}
\end{equation}

%\nocite{*}
\bibliographystyle{aipnum4-1}
\bibliography{ReferenceW}% Produces the bibliography via BibTeX.

\end{document}